\documentclass{aa}  

\usepackage{txfonts,epsfig,graphicx,url,twoopt}
\usepackage[breaklinks=true]{hyperref} 
\hypersetup{colorlinks=true,citecolor=blue,linkcolor=blue,urlcolor=black}
\graphicspath{{./figures/}}
\begin{document}

   \title{Planetesimal formation starts at the snow line}

   \subtitle{}

   \author{J. Dr{\k a}{\. z}kowska\inst{1}
          \and
          Y. Alibert\inst{2}
          }

   \institute{Institute for Computational Science, University of Zurich,
              Winterthurerstrasse 190, 8057 Zurich, Switzerland\\
              \email{joannad@physik.uzh.ch}
           \and
              Physikalisches Institut \& Center for Space and Habitability, University of Bern, 
              Sidlerstrasse 5, 3012 Bern, Switzerland\\
             }

   \date{Accepted on 28 September 2017}

 
  \abstract
   {Planetesimal formation stage represents a major gap in our understanding of the planet formation process. The late-stage planet accretion models typically make arbitrary assumptions about planetesimals and pebbles distribution while the dust evolution models predict that planetesimal formation is only possible at some orbital distances. }
   {We want to test the importance of water snow line for triggering formation of the first planetesimals during the gas-rich phase of protoplanetary disk, when cores of giant planets have to form.}
   {We connect prescriptions for gas disk evolution, dust growth and fragmentation, water ice evaporation and recondensation, as well as transport of both solids and water vapor, and planetesimal formation via streaming instability into a single, one-dimensional model for protoplanetary disk evolution.}
   {We find that processes taking place around the snow line facilitate planetesimal formation in two ways. First, due to the change of sticking properties between wet and dry aggregates, there is a "traffic jam" inside of the snow line that slows down the fall of solids onto the star. Second, ice evaporation and outward diffusion of water followed by its recondensation increases the abundance of icy pebbles that trigger planetesimal formation via streaming instability just outside of the snow line.}
   {Planetesimal formation is hindered by growth barriers and radial drift and thus requires particular conditions to take place. Snow line is a favorable location where planetesimal formation is possible for a wide range of conditions, but still not in every protoplanetary disk model. This process is particularly promoted in large, cool disks with low intrinsic turbulence and increased initial dust-to-gas ratio.}
   {}  

   \keywords{accretion, accretion disks -- 
                stars: circumstellar matter -- 
                protoplanetary disks -- 
                planet and satellites: formation -- 
                methods: numerical
               }

   \maketitle
%

\section{Introduction}

Our understanding of planet formation is severely limited by the fact that we cannot explain the connection between its early and late stages. As a consequence, models that deal with the late-stage planet accretion, when a planetary embryo grows to its final size and structure, typically use the same input for the radial distribution of gas and solids as the early-stage models dealing with dust growth and planetesimal formation. However, the latter models show that growing large bodies is not easy. This is because of the growth barriers: the dust growth is inhibited at centimeter sizes and some particular conditions are needed for the formation of larger, gravitationally bound planetesimals and planetary embryos. 

Probably the most widely accepted planetesimal formation scenario at the moment is the streaming instability \citep{2005ApJ...620..459Y, 2007Natur.448.1022J}. For sufficiently large pebbles and increased metallicity \citep{2010ApJ...722.1437B, 2014A&A...572A..78D, 2015A&A...579A..43C}, streaming instability leads to formation of dense filaments that become gravitationally unstable and collapse to planetesimals. This scenario allows us to bypass the growth barriers and form gravitationally bound object directly from pebbles. 

The streaming instability is typically simulated in local boxes due to the high computational cost of hydrodynamical simulations \citep{2011A&A...529A..62J, 2013MNRAS.434.1460K, 2016ApJ...822...55S}. Thus, the initial conditions are already set up in a way that streaming instability occurs. However these conditions do not necessarily happen in a realistic disk that starts its evolution with dust-to-gas ratio on the order of 1\%, which gets depleted because of removal of solids by the radial drift \citep{2010A&A...513A..79B,2012MNRAS.423..389H,2016A&A...586A..20K}. Pebble pile-ups may be necessary to allow for planet formation in the gas-rich phase of protoplanetary disk \citep{2016A&A...594A.105D, 2017MNRAS.467.1984G}, while disk dispersal via photoevaporation may allow for late planetesimal formation \citep{2017ApJ...839...16C}. Both processes may be needed to explain the existence of different planet types and debris belts, for instance in the Solar System. In this paper, we focus on the former mechanism, with the aim of triggering planetesimal formation early in the evolution of gas disk in order to allow sufficient time for the formation of gas-rich planets. 

The great dichotomy of the Solar System used to be commonly attributed to the jump change of condition around the snow line \citep{1988Icar...75..146S, 2000prpl.conf.1081W, 2015Icar..258..418M}, with water ice greatly enhancing the abundance of solids outside of this point. 
Such a rapid change of conditions may contribute to a pressure bump build-up that would halt the radial drift of solids thus facilitating planet formation \citep{2007ApJ...664L..55K, 2008A&A...487L...1B, 2013A&A...556A..37D}.
It was also proposed that the icy dust aggregates can pile-up \citep{2004ApJ...614..490C} or even significantly grow thanks to water vapor recondensation at the snow line \citep{2013A&A...552A.137R,  2015MNRAS.449.1084W}. Laboratory and numerical experiments dealing with collisional properties of dust aggregates concluded that the icy aggregates are significantly more sticky than silicate grains \citep{2011ApJ...737...36W, 2011Icar..214..717G} and can thus grow to larger sizes before they fragment, or even grow directly to planetesimal sizes if they are sufficiently porous \citep{2012ApJ...752..106O, 2013A&A...557L...4K}.

In this paper, we analyze how the presence of the snow line could trigger formation of the first gravitationally bound planetesimals at the very beginning of planet formation. We start our simulations with a smooth protoplanetary disk and let it evolve taking into account dust growth to pebbles, their drift and fragmentation, as well as ice evaporation and recondensation. In order to demonstrate the universality of our findings, we test our scenario in three different protoplanetary disk models. We conclude that the water component has an immense effect on the growth and redistribution of solids and leads to a pile-up of icy pebbles and planetesimal formation via streaming instability just outside of the snow line. Significant contribution to this pile-up comes from the change of sticking properties between icy and dry aggregates, an effect that was previously included in some of the models \citep{2010A&A...513A..79B, 2015ApJ...815L..15B, 2016ApJ...818..200E, 2017MNRAS.465.3865C} but not discussed directly in the context of planetesimal formation. 

This paper is organized as follows. We describe our numerical modeling approach and typical initial conditions in Sect.~\ref{sub:models} and present typical results and their dependence on input parameters in Sect.~\ref{sub:results}. We discuss the differences between our work and other published results as well as the implications of our findings in Sect.~\ref{sub:discussion}, and finally summarize our work in Sect.~\ref{sub:summary}.

\section{Model}\label{sub:models}

We implement one-dimensional protoplanetary disk model where we follow the radial distribution of solids and formation of planetesimals over one million years. We focus on the gas-rich phase of protoplanetary disk, before photoevaporation is efficient, thus we either include only viscous evolution or implement a static gas disk. The initial dust grains size is 1~$\mu$m at every orbital distance. We assume that the dust may be composed of water ice and silicates, and that the initial ice mass fraction outside of the snow line is 50\%. At the beginning of each simulation, the ice and water vapor are distributed across the disk such that the vertically integrated water-to-gas ratio is uniform and equal to 0.5 of the total metallicity. Solid ice is only present outside of the snow line, and water vapor is present inside of the snow line. The refractory dust component is present both outside and inside of the snow line. We follow dust growth to pebbles, fragmentation, and radial drift as well as ice evaporation and recondensation of water vapor.

Solids and water vapor evolution is governed by their interactions with the sub-Keplerian, turbulent gas. We treat the gas disk as a background for solids evolution and plug-in different gas disk models, as described in the following section.

\subsection{Gas disk models}

\begin{figure}
   \centering
   \includegraphics[width=0.95\hsize]{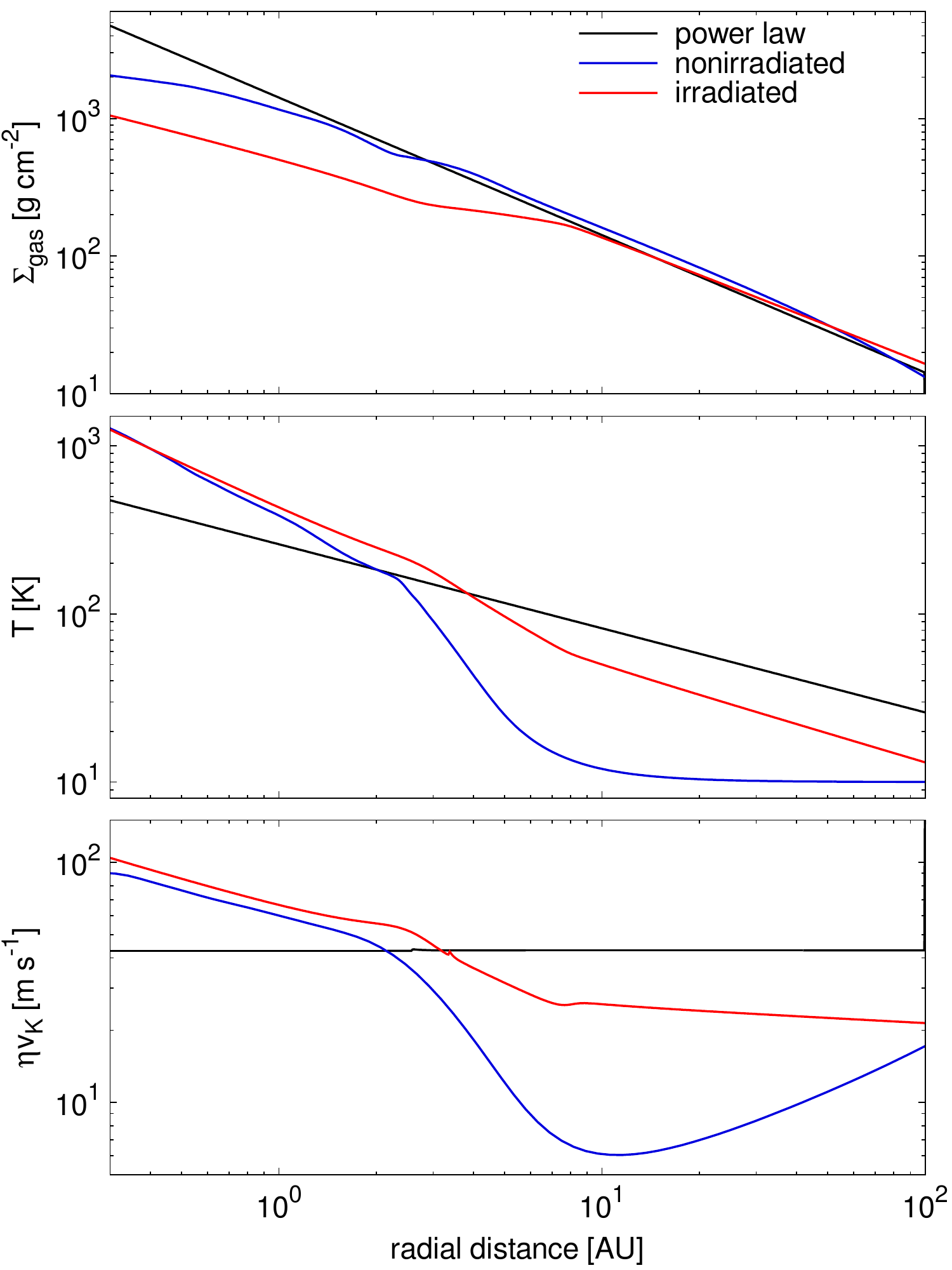}
      \caption{Comparison of the initial conditions for our protoplanetary disk models: the power-law disk (black), the non-irradiated disk (blue), and the irradiated disk (red). The panels show, from top to bottom, gas surface density $\Sigma_{\rm gas}$, temperature $T$, and the difference between gas and Keplerian rotation, which is equal to the maximum drift speed of dust pebbles $\eta v_{\rm K}$. Each of the disks has the total mass of 0.1~$M_{\odot}$. }
      \label{fig:disks}
\end{figure}

To test versatility of the planetesimal formation scenario discussed in this paper, we will use three different protoplanetary disk models. All of them have total gas mass of $0.1$~M$_{\odot}$ within 100~AU distance to the central star of $M_\star=1$~M$_{\odot}$. Their other properties are described in subsequent paragraphs. Comparison of the basic properties of these disks: surface density, temperature, and deviation from the Keplerian rotation, is displayed in Fig.~\ref{fig:disks}.

\subsubsection{Power-law disk}

Very simple protoplanetary disk models are commonly used in planet formation research. Here, we adopt one of them, with gas surface density profile set as
the straightforward function of the distance to the central star $r$
\begin{equation}
{\Sigma_{\rm g}} = 1400 \cdot \left(\frac{r}{\rm AU}\right)^{-1},
\end{equation}
and temperature profile is fixed to
\begin{equation}
{T} = 150 \cdot \left(\frac{r}{3~{\rm AU}}\right)^{-1/2}.
\end{equation}
For simplicity, as the model is already basic, we keep this disk static, however we include effects of gas accretion velocity $v_{\rm g}$ when calculating radial drift of solids. This velocity is estimated as 
\begin{equation}\label{vgas}
v_{\rm g} = \frac{3}{2}\frac{D_{\rm g}}{r},
\end{equation}
where turbulent gas diffusivity is equal to gas viscosity
\begin{equation}
D_{\rm g} = \nu = \alpha_{\rm v} \frac{c_{\rm s}^2}{\Omega_{\rm K}},
\end{equation}
which is calculated based on the standard $\alpha$-accretion model \citep{1973A&A....24..337S} with $\alpha_{\rm v}$ being the dimensionless parameter describing the efficiency of angular momentum transfer. We typically use $\alpha_{\rm v}=10^{-3}$ in this paper. The $\Omega_{\rm K}$ is the Keplerian orbital frequency and the gas sound speed $c_{\rm s}$ is calculated as
\begin{equation}
c_{\rm s} = \sqrt{\frac{k_{\rm B}T}{\mu}},
\end{equation}
where $k_{\rm B}$ is the Boltzmann constant and $\mu$ is the mean molecular weight of gas. When calculating $\mu$, we include the contribution of water vapor:
\begin{equation}
\mu = \left(\Sigma_{\rm g}+\Sigma_{\rm H_2O}\right)\cdot\left({\frac{\Sigma_{\rm g}}{\mu_{\rm g}}+\frac{\Sigma_{\rm H_2O}}{\mu_{\rm H_2O}}}\right)^{-1},
\end{equation}
which increases the molecular weight and thus decreases the sound speed in the inner part of the protoplanetary disk. We take $\mu_{\rm g}=2.34$~m$_{\rm p}$ and $\mu_{\rm H_2O}=18$~m$_{\rm p}$, with m$_{\rm p}$ denoting the proton mass. The effect of variable $\mu$ is visible in the bottom panel of Fig.~\ref{fig:disks} as a small jump of the maximum drift speed (equal to the difference between the gas and Keplerian rotation) around the snow line location. The maximum radial drift speed is roughly constant in the power-law disk model, with $\eta v_{\rm K} \approx 40$~m~s$^{-1}$.

\subsubsection{Non-irradiated disk}

We compare the simple static power-law disk model to a more complex, viscously evolving circumstellar disk model described in \citet{2005A&A...434..343A, 2013A&A...558A.109A}, which was also used in our previous work presented in \citet{2016A&A...594A.105D}. In this model, the vertical structure of the disk is first computed for every distance to the star by solving the hydrostatic equation, the energy conservation equation, and the radiative diffusion equation (energy is assumed to be transported by radiation). The solution of the vertical structure equations gives the thermodynamical quantities (temperature, pressure, density, but also disk scale height) as a function of the gas surface density. The same calculation also gives the vertically averaged viscosity, which is again computed in the framework of the $\alpha$ formalism. We finally solve the diffusion equation, the viscosity being the one derived from the vertical structure calculation, in order to compute the time evolution of the gas surface density. 

As can be seen in the middle panel of Fig.~\ref{fig:disks}, this disk is very cold in its outer part, as the temperature drops down to 10~K outside of 10~AU. Inside of the 10~AU, the temperature profile is steeper than the one implemented in the power-law disk and reaches over 1000~K at the inner edge of the disk. This steeper temperature profile leads to a flatter surface density in the inner part of the disk, while it is very similar to the power-law disk in the outer part. In the inner part of the disk, where the temperature profile is steep, the maximum radial drift speed is higher than for the power-law disk, but it drops in the outer disk, forming a wide minimum around 10~AU (see the bottom panel of Fig.~\ref{fig:disks}).  

\subsubsection{Irradiated disk}

To test the effect of stellar irradiation, we implement a simple analytical model proposed by \citet{2015A&A...575A..28B} (B15 in the following), which was designed to fit 2-D radiative hydrodynamic simulations of protoplanetary disks. 
In this model the disk evolves with time and the accretion rate $\dot{M}$ decreases as follows \citep{1998ApJ...495..385H}
\begin{equation}
\log \left( {\dot{M} \over M_\odot / {\rm yr}} \right) = -8.00 -1.40 \cdot \log \left( {t + 10^5 {\rm yr} \over 10^6 {\rm yr}} \right)
\end{equation}
where $t$ is the evolution time and $\dot{M}$ is related to the viscosity and the gas surface density ${\Sigma_{\rm g}}$ via
\begin{equation}
\dot{M} = 3 \pi \nu {\Sigma_{\rm g}} = 3 \pi \alpha_{\rm v} H_{\rm g}^2 \Omega_{\rm K} {\Sigma_{\rm g}},
\end{equation}
where $H_{\rm g}$ the scale height of the disk, and $\Omega_{\rm K}$ the Keplerian frequency. In this disk model, we use $\alpha_{\rm v} = 0.0054$ following B15. We note that $\alpha_{\rm v}$ is only used in B15 as a heating parameter and not to evolve the disk viscously as it is in the case of the non-irradiated disk model. 
For a given $\dot{M}$, all quantities of the disk can be derived, except for the temperature profile. In order 
to compute the latter, we use the formulas presented in Appendix~A of B15.

Once the temperature is determined, it can be linked to the disk aspect ratio via
\begin{equation}
T = \left( { H_{\rm g} \over r } \right)^2 {G M_\star \over r } {\mu \over \mathcal{R}}
\end{equation}
where $G$ is the gravitational constant, $r$ the location in the disk, $\mathcal{R}$ is the gas constant and $\mu$ is the mean molecular weight. 

As can be seen in Fig.~\ref{fig:disks}, the temperature in this disk is significantly higher than in the non-irradiated disk, particularly in its outer part, thus the surface density flattening is also more pronounced. Also, the maximum drift speed is higher than in the non-irradiated disk, but still lower than the power-law disk in the outer parts. 

\subsection{Evolution of solids}

We follow the solids surface density $\Sigma_{\rm d}$ evolution by solving the advection-diffusion equation
\begin{equation}\label{advdiff}
\frac{\partial \Sigma_{\rm d}}{\partial t} + \frac{1}{r} \frac{\partial}{\partial r}\left[r\left(\Sigma_{\rm d}\bar{v}-D_{\rm g}\Sigma_{\rm g}\frac{\partial}{\partial r}\left(\frac{\Sigma_{\rm d}}{\Sigma_{\rm g}}\right)\right)\right] = 0,
\end{equation}
where $\bar{v}$ is the mass weighed average radial velocity of solids, which encompasses information about their size.

Dust aggregates sizes are modeled using method based on the two-population algorithm proposed by \citet{2012A&A...539A.148B}. The logic behind this algorithm is to reduce computational intensity by not solving for the dust coagulation directly but rather predict its outcome based on more complete models. Thus, the dust size distribution is set in each radial grid cell depending on the dominating process: coagulation-fragmentation equilibrium or radial drift. We waive the description of further details of this model as they are thoroughly outlined in the original work as well as in our previous paper \citep{2016A&A...594A.105D}. Importantly, the outcome of this algorithm is the size, or Stokes number, distribution at every radial distance. The Stokes number informs us about the interaction between the solids grain and gas, and is connected to the grain size $a$:
\begin{equation}\label{stnr}
{\rm{St}} = \frac{\pi}{2}\frac{a\rho_{\bullet}}{\Sigma_{\rm{g}}},
\end{equation}
where $\rho_{\bullet}$ is the aggregate internal density. Eq.~(\ref{stnr}) is derived under an assumption that the solids are in the Epstein drag regime, which means that their sizes do not exceed the mean free path in the gas, which is true for all our models. The internal density of aggregates is calculated based on their composition:
\begin{equation}\label{rhop}
\rho_{\bullet} = \left(m_{\rm sil}+m_{\rm ice}\right)\cdot\left({\frac{m_{\rm sil}}{\rho_{\bullet,{\rm sil}}}+\frac{m_{\rm ice}}{\rho_{\bullet,{\rm ice}}}}\right)^{-1},
\end{equation}
where $m_{\rm sil}$ and $m_{\rm ice}$ are mass of silicate and mass of ice contained within the aggregate.
A pure water ice aggregate would have $\rho_{\bullet,{\rm ice}}=1$~g~cm$^{-3}$ and a silicate aggregate $\rho_{\bullet,{\rm sil}}=3$~g~cm$^{-3}$.

When calculating the advection speed of solids $\bar{v}$, we take into account both radial drift caused by the interaction with the sub-Keplerian gas and by gas accretion flow:
\begin{equation}\label{vbar}
\bar{v} = -\frac{2\eta v_{\rm K} \bar{\rm St} + v_{\rm g} \left(1+\epsilon\right)}{{\bar{\rm St}}^2 + \left(1+\epsilon\right)^2} 
\end{equation}
where $\eta v_{\rm K}$ is the maximum radial drift speed driven by the radial gas pressure gradient:
\begin{equation}
\eta v_{\rm K} = -\frac{1}{2}\frac{c_{\rm s}^2}{v_{\rm K}}\frac{\partial \log P}{\partial \log r},
\end{equation}
where $v_{\rm K}$ is the Keplerian velocity of gas and $P$ is pressure calculated taking into account contributions both from nebular hydrogen and helium gas and water vapor. $\bar{\rm St}$ is the mass weighted average Stokes number of solids at given radial distance, which is calculated from the size distribution returned by the above-mentioned algorithm. The gas accretion velocity $v_{\rm g}$ is calculated as in Eq.~(\ref{vgas}). We take into account the collective drift effect, which means that the drift velocity decreases as the solids-to-gas ratio increases. As most of the pebbles are settled to the midplane, we implement the midplane solids-to-gas ratio $\epsilon=\rho_{\rm d}/\rho_{\rm g}$ in Eq.~(\ref{vbar}). This equation is equivalent to the one used by \citet{2016A&A...596L...3I} and \citet{2017A&A...602A..21S}. 

\subsubsection{Fragmentation threshold}\label{sub:vf}

Evolution of solids is dominated by the radial drift and by fragmentation that may stop the growth as the impact speeds increase with the Stokes number of grains (for ${\rm St}<1$). The maximum aggregate size that can be obtained before fragmentation kicks in is sensitively dependent on the fragmentation threshold velocity $v_{\rm f}$ \citep[see][]{2012A&A...539A.148B}:
\begin{equation}\label{afrag}
a_{\rm{frag}} \propto \frac{v_{\rm{f}}^2}{\alpha_{\rm t} c_{\rm{s}}^2},
\end{equation}
so the choice of $v_{\rm{f}}$ value is in fact very important to the model outcome.  $\alpha_{\rm t}$ is the midplane turbulence strength parameter that regulates impact speeds of pebbles and their settling. We purposely distinguish $\alpha_{\rm t}$ from $\alpha_{\rm v}$, the efficiency of angular momentum transport via turbulent viscosity that is used in the gas disk models. This is motivated by the fact that in many recent protoplanetary disk models the angular momentum transfer is not necessarily driven by turbulence anymore, and even if it is, a quiescent midplane layer is often formed \citep{2013ApJ...765..114D, 2014prpl.conf..411T, 2016ApJ...821...80B}. In most of our runs we assume $\alpha_{\rm t} = 10^{-3}$, but we discuss the impact of this value in Sect.~\ref{sub:alpha}.

Laboratory experiments estimated threshold velocities of around 1~m~s$^{-1}$ for the onset of fragmentation of silicate dust aggregates \citep[see e.g.][]{2010A&A...513A..56G}. It is commonly accepted that aggregates containing water ice fragment at higher velocities, as their surface energies are about 10 times higher than those of silicates \citep{2011ApJ...737...36W, 2011Icar..214..717G, 2014MNRAS.437..690A, 2015ApJ...798...34G}. Thus, we set the fragmentation threshold velocity according to aggregates composition. For dry aggregates we set $v_{\rm{f,in}} = 1$~m~s$^{-1}$ and for aggregates containing more than 1\% of water ice $v_{\rm{f,out}} = 10$~m~s$^{-1}$. This threshold amount of ice above which we consider our aggregates to be more sticky is arbitrary, but we tested that the exact value does not make much difference to the results of our models as the ice-to-dust ratio drops very rapidly across the snow line.

\subsubsection{Evaporation and recondensation}\label{sub:evapcond}

We take water ice evaporation and recondensation of water vapor into account following the treatment proposed by \citet{2006Icar..181..178C}. We trace the transport of solid ice that is incorporated into aggregates and thus migrates through the disk much faster than the gas, and the water vapor that moves at the same speed as the gas. To follow the evolution of water vapor surface density $\Sigma_{\rm vap}$, we solve the following transport equation:
\begin{equation}\label{vapdiff}
\frac{\partial \Sigma_{\rm vap}}{\partial t} + \frac{1}{r} \frac{\partial}{\partial r}\left[r\left(\Sigma_{\rm vap}v_{\rm g}-D_{\rm g}\Sigma_{\rm g}\frac{\partial}{\partial r}\left(\frac{\Sigma_{\rm vap}}{\Sigma_{\rm g}}\right)\right)\right] = 0,
\end{equation}
where $v_{\rm g}$ is the gas velocity. Eq.~(\ref{vapdiff}) is analogous to Eq.~(\ref{advdiff}) used to follow the transport of solids.

We assume that all the dust grains at a given orbital distance (i.e.~in a given radial bin) have the same composition (i.e.~ice mass ratio), which means that water is added to and removed from the dust with constant $dm/m$ during recondensation and evaporation. This is consistent with instantaneous redistribution of ice component because of coagulation and fragmentation, which happens when the collisional timescale
\begin{equation}\label{tgrowth}
\tau_{\rm growth} = \frac{a}{\dot{a}} \approx \frac{1}{Z\cdot\Omega_{\rm K}}
\end{equation}
\citep[see][]{2012A&A...539A.148B} is shorter than the radial drift timescale
\begin{equation}\label{tdrift}
\tau_{\rm drift} = \frac{r}{|v_{\rm r,d}|}.
\end{equation}
In our runs, this is indeed true around the snow line, where nominal timescales are $\tau_{\rm growth}\approx10^3$~years and $\tau_{\rm drift}\approx10^4$~years. When the pebble pile-up is formed around the snow line, the vertically integrated dust-to-gas ratio $Z$ increases making the growth timescale even shorter, and the radial drift speed $v_{\rm r,d}$ decreases (because of the collective drift effect) making the drift timescale even longer.

At every timestep, we calculate the equilibrium pressure, which is given by the Clausius-Clapeyron equation:
\begin{equation}
P_{\rm eq} = P_{{\rm eq},0}\cdot\exp\left({-\frac{A}{T}}\right),
\end{equation}
where the constants $P_{{\rm eq},0}=1.14\cdot10^{13}$~g~cm$^{-1}$~s$^{-2}$ and $A=6062$~K are taken from \citet{1991Icar...90..319L}. We compare the value of $P_{\rm eq}$ to the water vapor pressure
\begin{equation}
P_{\rm vap} = \frac{\Sigma_{\rm vap}}{\sqrt{2\pi}H_{\rm g}}\cdot\frac{k_{\rm B}T}{\mu_{\rm H_2O}},
\end{equation}
where $\Sigma_{\rm vap}$ is the water vapor surface density and $H_{\rm g} = c_{\rm s}/\Omega_{\rm K}$ is the gas scale-height. 

If $P_{\rm vap} < P_{\rm eq}$, evaporation takes place and the surface density of ice decreases by
\begin{equation}\label{evap}
\Delta{\Sigma}_{\rm ice} = \min\left(\sqrt{\frac{8\pi\mu_{\rm H_2O}}{k_{\rm B}T}}\cdot\frac{\bar{a}^2}{\bar{m}}\cdot P_{\rm eq}\cdot\Sigma_{\rm ice}\cdot\Delta{t},{\Sigma}_{\rm ice}\right)
\end{equation}
where $\bar{a}$ and $\bar{m}$ are the average size and mass of pebbles aggregates respectively, and $\Delta{t}$ is the time-step. The material removed from solid ice phase is added to the vapor reservoir. 

If $P_{\rm vap} > P_{\rm eq}$, the water vapor condenses onto grains and the surface density of ice is increased by
\begin{equation}\label{recon}
\Delta{\Sigma}_{\rm ice} = \min\left(2H_{\rm g}\cdot\frac{\mu_{\rm g}}{k_{\rm B}T}\cdot\left(P_{\rm vap} - P_{\rm eq}\right),\Sigma_{\rm vap}\right),
\end{equation}
which essentially means that all the excess vapor is added to the solid phase, such that the vapor pressure drops to the equilibrium pressure. The surface density added to the ice is subsequently removed from the vapor supply. 

In the original work of \citet{2006Icar..181..178C} there was a distinction between populations of dust and migrators (pebbles). Evaporation was considered to happen both from dust and migrators, while condensation was only happening on dust, thus it is assumed to be instantaneous. In our model, we do not explicitly make this distinction between dust and pebbles, but we assume that there is a continuous size distribution in every radial cell, which around the snow line is set by coagulation-fragmentation equilibrium. However, we do not model the evaporation and condensation on every size bin but take into account the surface-weighted average size $\bar{a}$. Condensation on grains larger than micron-sized should take some time because they have smaller surface area available. In reality however, vapor is mostly condensing on the smallest grains that have the most surface area available \citep[see e.g.][]{2017A&A...600A.140S}. Since the snow line region is in the fragmentation dominated regime (see Fig.~\ref{fig:sizes}), the small grains should be constantly replenished, so we keep the assumption of instantaneous recondensation. In practice, the same assumption might have been made for evaporation, as it is very fast (a centimeter sized pebble crossing the snow line would lose its ice content within $\tau_{\rm{evap}}\approx1$~year) in all of the runs presented in this paper. 

\subsubsection{Planetesimal formation via streaming instability}\label{sub:pfmodel}

With our one-dimensional model, we cannot resolve streaming instability that would locally condense pebbles into dense filaments, which would then gradually collapse to form planetesimals. To include the possibility of planetesimal formation via this process, we use the same approach as we did in \citet{2016A&A...594A.105D}. At every time-step and in every radial bin, we check if the midplane density of pebbles exceeds unity. With the turbulence level parameter $\alpha_{\rm t} \geq 10^{-4}$ that we use in this paper, this condition is always stronger than the critical metallicity conditions proposed for laminar disk by \citet{2014A&A...572A..78D} and \citet{2015A&A...579A..43C} (recently \citet{2017ApJ...839...16C} arrived at the same conclusion). 

If the criterion for planetesimal formation via streaming instability is fulfilled, namely $\rho_{\rm d}({\rm St}>10^{-2})/\rho_{\rm g} > 1$, part of the surface density of pebbles is transferred to planetesimals:
\begin{equation}
\dot{\Sigma}_{\rm{plts}} = \zeta\cdot\Sigma_{\rm d}({\rm St}>10^{-2})\cdot\Omega_{\rm K},
\end{equation}
with the efficiency of $\zeta = 10^{-3}$, that is motivated by numerical models presented by \citet{2016ApJ...822...55S} and for which \citet{2016A&A...594A.105D} found convergence of the amount of planetesimal formed, i.e.~the amount of planetesimal would not change significantly for higher $\zeta$ values. 

\subsection{Model assumptions}

Our algorithm is limited by several assumptions, which we list here for clarity.
\begin{itemize}
\item{We consider one-dimensional, locally isothermal disk models and focus on the evolution of their midplane, where pebbles and planetesimals reside. Thus, we only consider the radial snow line and neglect effects associated with the atmospheric snow line.}
\item{Grain sizes are set by either the coagulation-fragmentation or the growth-drift equilibrium. We do not consider the impact of evaporation and condensation on aggregate sizes. This is equivalent with assuming that the ice added during recondensation is quickly redistributed by coagulation and fragmentation. Any aggregates that would increase their size over the maximum are immediately fragmented or removed by radial drift.}
\item{During one time-step and at a given orbital distance, all the aggregates are either in evaporation or in condensation mode. We do not consider grain curvature effects that could switch between those effects from grain to grain.  }
\item{We treat all grains as compact spheres and neglect effects of porosity.}
\item{We assume that vertical structure of solids is always in equilibrium between settling and turbulent mixing, which leads to the dust scale-height derived by \citet{1995Icar..114..237D}
\begin{equation}\label{hdust}
H_{\rm d} = H_{\rm g} \sqrt{\frac{\alpha_{\rm t}}{\alpha_{\rm t} + {\rm St}}},
\end{equation}
and that the water vapor is instantly mixed up to gas scale-height $H_{\rm g}$, even though it is released by pebbles with $H_{\rm d} < H_{\rm g}$. Recently, \citet{2016ApJ...833..285K} showed that the Eq.~(\ref{hdust}) breaks at high dust-to-gas ratios, when the collisional evolution is faster than the vertical redistribution.}
\item{We assume that minimum pebble size necessary to trigger planetesimal formation via streaming instability corresponds to $\rm{St}=10^{-2}$ following \citet{2010ApJ...722.1437B} and \citet{2014A&A...572A..78D}. More recently, \citet{2015A&A...579A..43C} and \citet{2016arXiv161107014Y} suggested that streaming instability is also possible for smaller grains, albeit at higher metallicity. However, we find that for our assumed turbulence strength, the smaller grains do not settle the the midplane efficiently (see Eq.~(\ref{hdust})), so their contribution to planetesimal formation would not be possible anyway in our models.}
\item{The structure of our gas disks is independent on solids evolution. This looses validity for high dust-to-gas ratio, when $\Sigma_{\rm d} \approx \Sigma_{\rm g}$, which happens when the pile-up forms in our models. Recently, \citet{2017MNRAS.467.1984G} and \citet{2017ApJ...844..142K} showed that including effects of dust backreaction on gas disk promotes formation and sustainability of dust pile-ups.}
\end{itemize}

\section{Results}\label{sub:results}

\subsection{Traffic jam inside and pebble pile-up outside of the snow line}

\begin{figure}
   \centering
   \includegraphics[width=0.9\hsize]{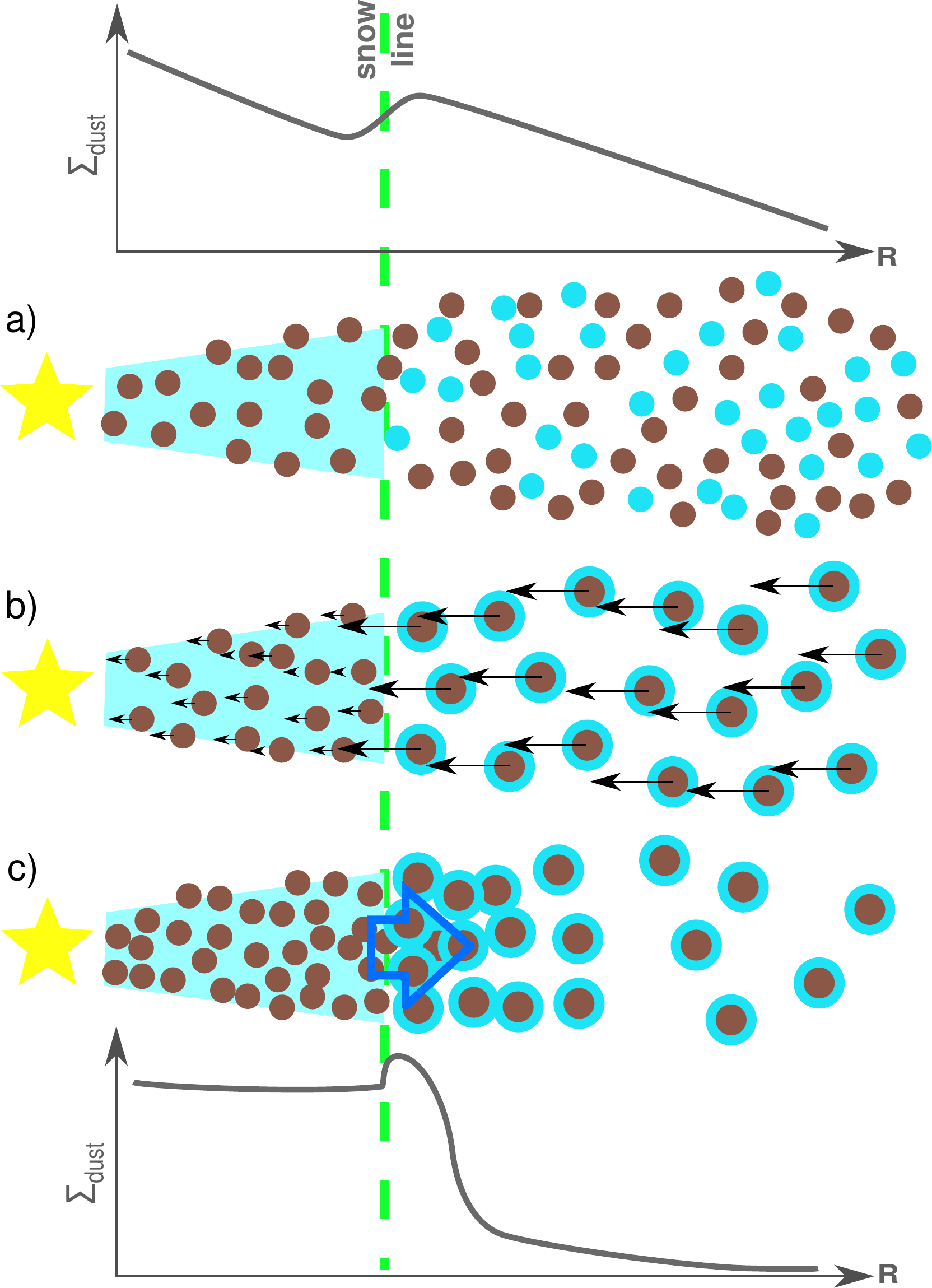}
      \caption{This sketch explains processes facilitating formation of the snow line pile-up: a)~In the initial condition, ice presence increases solids density outside of the snow line; b)~Coagulation is more efficient for aggregates that incorporate water ice. Thus, solids grow to larger sizes and drift faster outside of the snow line. The quick drift results in an efficient delivery of the embedded refractory material, which do not drift rapidly, causing "traffic jam" and increase of dust concentration in the inner disk; c)~The outward diffusion and recondensation of water vapor locally enhances abundance of solids just outside of the snow line, contributing to the pile-up of icy pebbles.}
      \label{fig:sketch}
\end{figure}

The common outcome of all the runs, independently of the underlying gas disk model, is that the highest solids-to-gas ratios are obtained in the region directly outside of the snow line. A general pattern for formation of this snow line pile-up is explained in the Fig.~\ref{fig:sketch}.

The dust-to-gas ratio in the inner parts of the disk is enhanced because of the general pattern of dust transport by radial drift that shifts mass inwards. As explained by \citet{2012A&A...539A.148B}, if the maximum size of dust grains is regulated by fragmentation, the surface density of solids becomes proportional to $r^{-1.5}$. As the gas surface density is shallower, this redistribution leads to depletion of the outer disk and increase of solids-to-gas ratio in the inner disk. As demonstrated by \citet{2016A&A...594A.105D}, the magnitude of this increase may already be sufficient to trigger planetesimal formation at the inner edge of the disk at a timescale of $\sim10^5$~years. However, in the present work this picture is complicated by the difference in fragmentation speeds of aggregates outside and inside of the snow line.

Refractory aggregates fragment at lower impact velocities and thus reach sizes that are two orders of magnitude smaller than the icy aggregates (see Eq.~(\ref{afrag}) and Fig.~\ref{fig:sizes}). As the drift velocity decreases with decreasing Stokes number, dust is retained in the inner disk causing sort of "traffic jam" effect. The dry dust aggregates inside of the snow line are small and well-coupled to the gas, so they undergo significant diffusion and do not form any local pile-up, unlike the icy pebbles outside of the snow line. The outward diffusion of water vapor and subsequent recondensation causes further increase in surface density of the icy pebbles just outside of the snow line. These pebbles are large enough to trigger streaming instability and form planetesimals. The exact location, extent, and mass of resulting planetesimal annulus depends on applied disk model, as discussed in Sect.~\ref{sub:pf}. 

\subsection{Fiducial simulation}

\begin{figure}
   \centering
   \includegraphics[width=0.95\hsize]{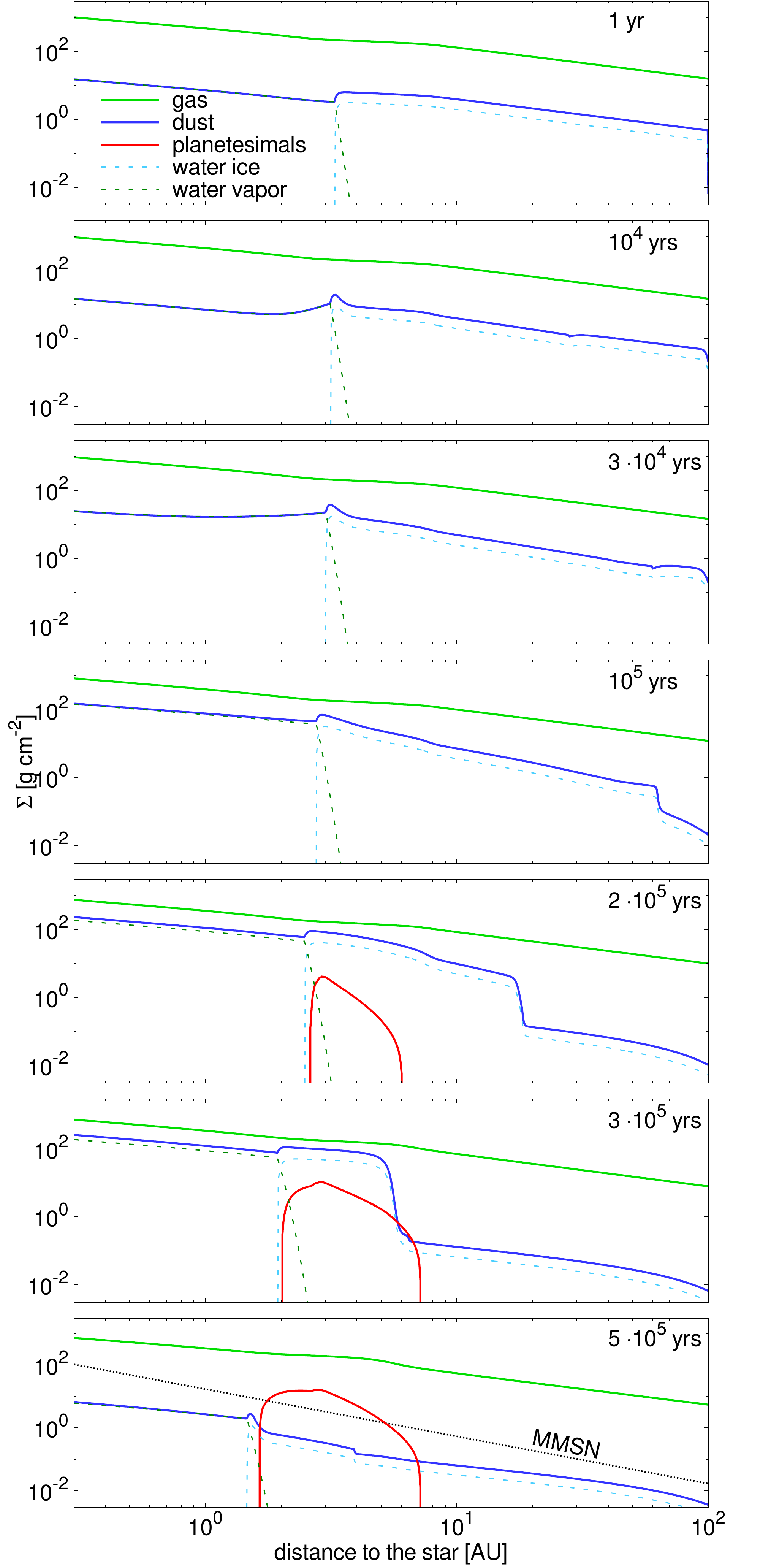}
      \caption{Panels show time evolution of surface density of gas, solids, planetesimals, water ice, and water vapor in the irradiated disk model with initial dust-to-gas ratio of $Z = 0.03$. In the bottom panel, surface density of solids in the classical minimum mass solar nebula model (MMSN) of \citet{1977Ap&SS..51..153W} is displayed for reference. Corresponding animation is available at {\url{http://www.ics.uzh.ch/\~joannad/snowline.mp4}.}}
      \label{fig:7panels}
\end{figure}

Figure~\ref{fig:7panels} shows the evolution of the gas and dust surface density, including ice and water vapor as well as the forming planetesimals in the irradiated disk model with initial dust-to-gas ratio of $Z = 0.03$, which we will refer to as the fiducial simulation later in the paper.

As can be seen in the uppermost panel of Fig.~\ref{fig:7panels}, presence of the solid ice increases the initial surface density in the outer part of the disk.
The predominant effect shaping the evolution of solids is their redistribution driven by growth and radial drift, which shifts mass inwards causing depletion of the outer parts and solids-to-gas ratio increase in the inner parts of the disk. 
Initially, the evolution outside of the snow line is dominated by fragmentation of the icy aggregates and thus the dust surface density is evolving toward $\Sigma_{\rm d}\propto r^{-1.5}$, as discussed in previous section. After $10^5$~years of evolution, as the outer disk gets depleted, it becomes dominated by the radial drift and the surface density profile at the outer edge becomes more shallow again, as it evolves to $\Sigma_{\rm d}\propto r^{-0.75}$ \citep{2012A&A...539A.148B}.

This picture is complicated here by ice evaporation and recondensation and different sticking properties of wet and dry aggregates. Keeping the grains inside of the snow line small, slows down their removal and keeps the dust-to-gas ratio in the inner disk high. The surface density of dust inside of the snow line keeps the same profile as the gas surface density for most of the time, because the aggregates are so small that their evolution is dominated by gas viscosity rather than radial drift (see Fig.~\ref{fig:sizes}). Ice evaporation leads to a jump in the surface density at the snow line. Recondensation of water vapor additionally bumps pebbles-to-gas ratio just outside of the snow line.
The combined action of the the "traffic jam" inside and pile-up outside of the snow line, which spreads outwards thanks to the collective drift effect (as the drift velocity decreases with increasing solids-to-gas ratio, see Eq.~(\ref{vbar})), leads to the conditions allowing for planetesimal formation via streaming instability.

The planetesimals start to appear after $2\cdot10^5$~years of evolution and their formation last for another $\sim2\cdot10^5$~years. During this time, the snow line, which marks the inner edge of planetesimal formation region, moves inwards as the disk cools down. At the same time, the planetesimal formation region slightly spreads outwards because of the collective drift effect. About 20~M$_{\oplus}$ of planetesimals are produced in this model. After $4\cdot10^5$~years, the inward flux of pebbles is not sufficient to supply the pile-up anymore and the surface density of solids quickly drops, terminating the planetesimal formation phase. 

\subsubsection{Pebble sizes}

\begin{figure}
   \centering
   \includegraphics[width=0.95\hsize]{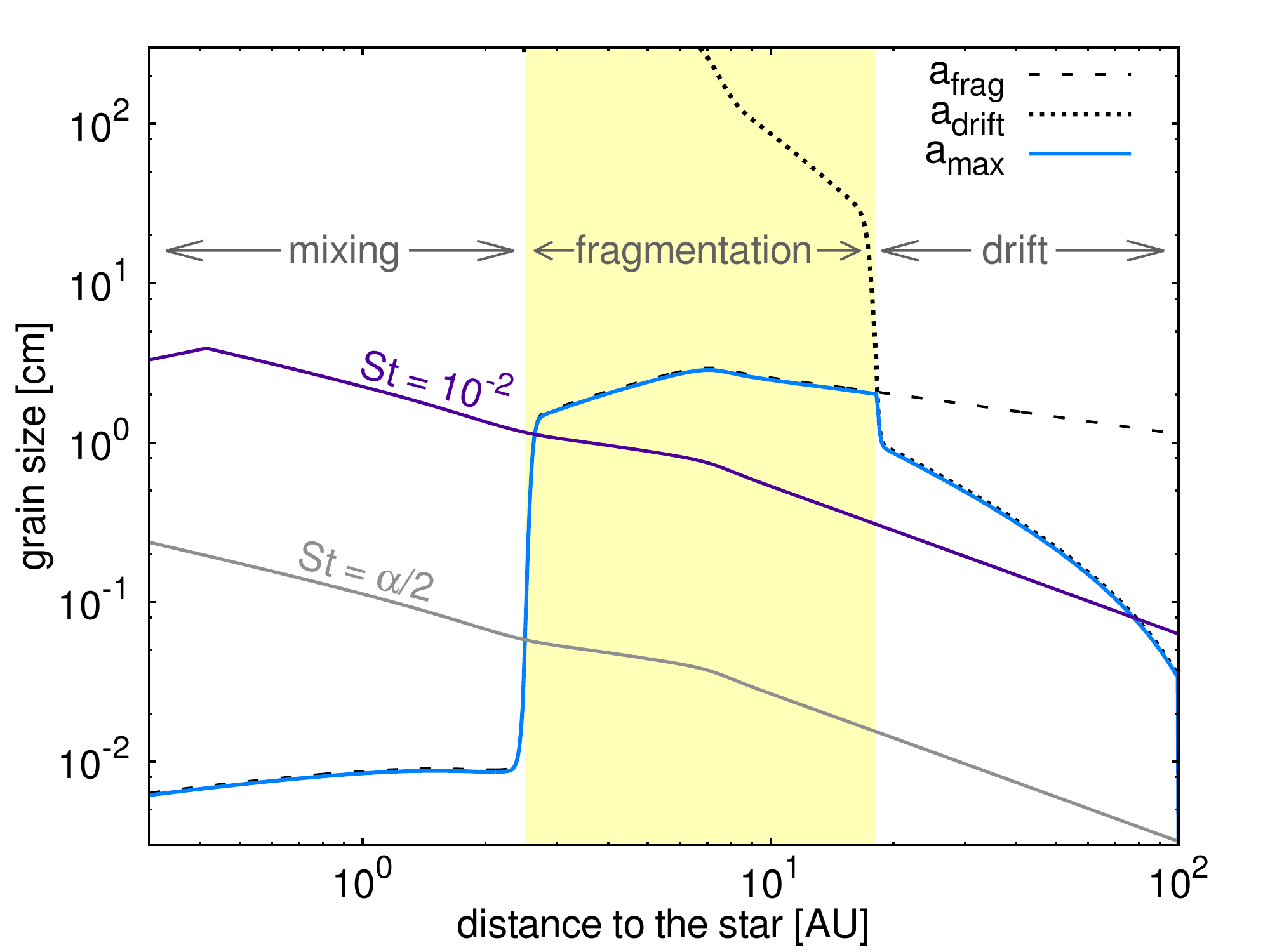}
      \caption{Maximum pebble size ($a_{\rm{max}}$) as a function of the radial distance after $2\cdot10^5$~years of evolution obtained in our fiducial run. In the outer part of the disk, this size is limited by the radial drift ($a_{\mathrm{drift}} < a_{\mathrm{frag}}$), while in the inner part of the disk it is limited by fragmentation. Aggregates that are large enough to participate in planetesimal formation via streaming instability (${\rm St}>10^{-2}$, purple solid line) are only present outside of the snow line. Inside of the snow line, aggregates size corresponds to ${\rm St}<\alpha_{\rm v}/2$ (gray solid line), which means their transport is dominated by viscosity rather than radial drift.}
      \label{fig:sizes}
\end{figure}

Figure \ref{fig:sizes} shows pebble sizes obtained from our simplified growth and fragmentation treatment after $2\cdot10^5$~years of evolution (during the period of planetesimal formation) in our fiducial simulation. These effective sizes are very similar in other runs.

The dust growth follows pattern described by \citet{2012A&A...539A.148B} and \citet{2016A&A...594A.105D}, with the inner disk being dominated by fragmentation driven by turbulence and the outer disk being gradually depleted by the radial drift before the particles have time to grow to the fragmentation limit. The modification of fragmentation velocity described in Sect.~\ref{sub:vf} introduces the rapid change of pebble size around the snow line. The icy pebbles outside of the snow line grow to sizes of several centimeters, corresponding to Stokes numbers of ${\rm St}>10^{-2}$, which allows for planetesimal formation. The dry aggregates inside of the snow line only grow to sub-millimeter sizes,  corresponding to ${\rm St}<\alpha_{\rm v}/2$, which means that they are well-coupled to the gas and follow its viscous evolution, or are in so-called mixing regime, as discussed by \citet{2012A&A...539A.148B}. Thus, these small grains drift at much lower speed than the icy pebbles, which contributes to enhancement of dust-to-gas ratio inside of the snow line and the retention of icy pebble pile-up outside of the snow line thanks to the outward diffusion of small grains.  

\subsubsection{Which is more important: traffic jam or recondensation?}\label{sub:what}

\begin{figure}
   \centering
   \includegraphics[width=0.95\hsize]{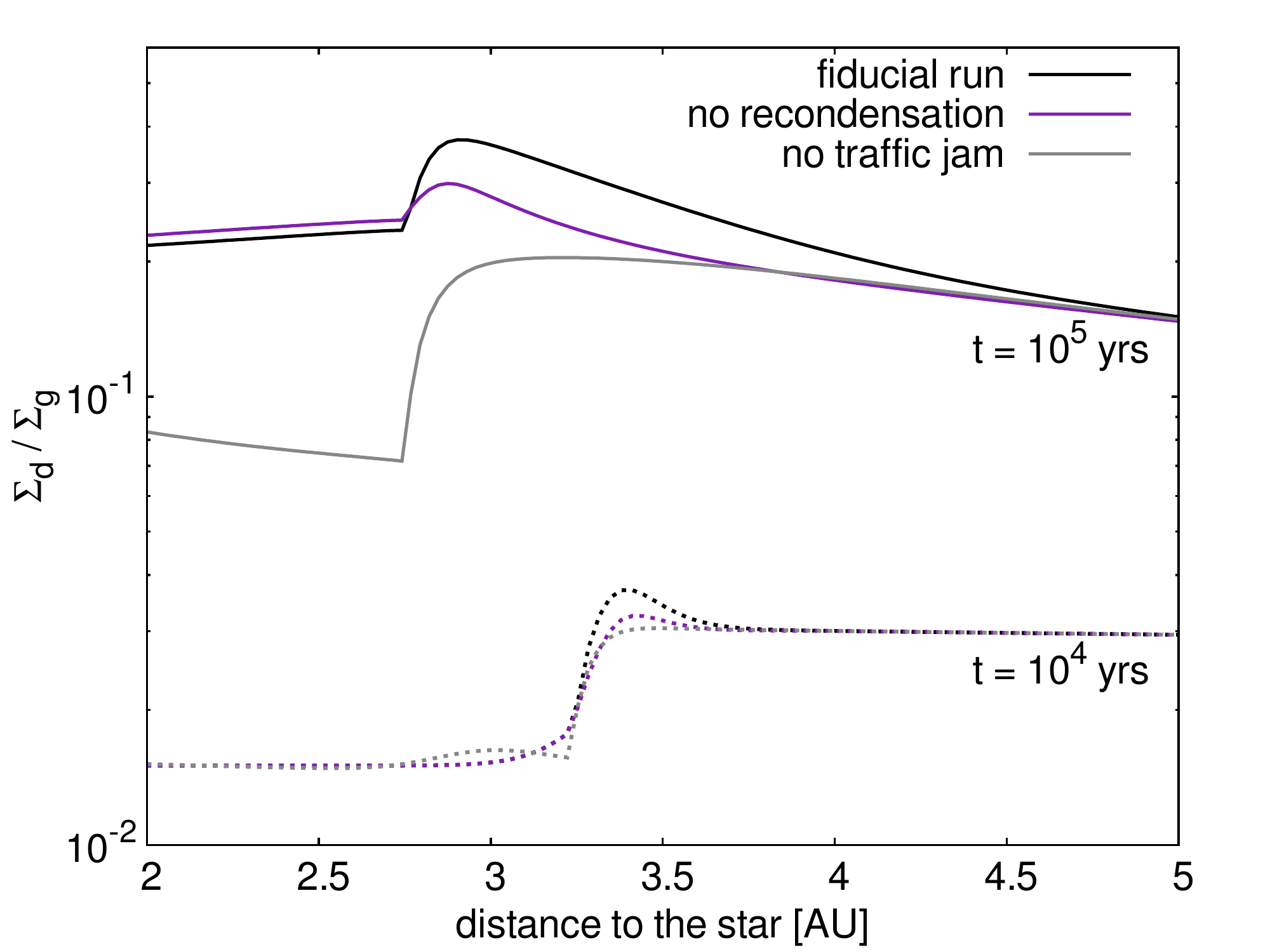}
      \caption{Vertically integrated solids-to-gas ratio around the snow line after $t=10^4$~yrs (dotted lines) and $t=10^5$~yrs (solid lines) of evolution for our fiducial run (black lines) and analogous runs without the water recondensation (purple lines) and without the difference between sticking properties of wet and dry aggregates (gray lines).}
      \label{fig:what}
\end{figure}

An obvious question arises: which of the two processes facilitating planetesimal formation at the snow line is more important? To answer this dilemma, we performed additional models with setup identical as in the fiducial run, but with one of the mechanisms switched off. Fig.~\ref{fig:what} compares solids-to-gas ratio around the snow line obtained in our fiducial run, the same run without recondensation of water vapor (but still including ice evaporation), and analogous run with fragmentation velocity for both wet and dry aggregates being equal to $v_{\rm f}=10$~m~s$^{-1}$, designed to get rid of the "traffic jam" effect inside of the snow line.

As can be seen in Fig.~\ref{fig:what}, the difference between sticking properties of wet and dry aggregates is the dominant process facilitating formation of the solids-to-gas ratio enhancement at the snow line at every evolutionary stage of the disk. Switching the recondensation off decreases the amplitude of the solids-to-gas ratio enhancement at the snow line only by $\sim20\%$. On the other hand, letting the dry aggregates grow to the same sizes as the icy pebbles leads to much more dramatic decrease of the solids-to-gas ratio in the bump. If the dry aggregates grow to pebble sizes, the solids-to-gas ratio falls more rapidly across the snow line and stays at lower levels inside of it. The surface density outside of the snow line reaches lower values and the pile-up vanishes more quickly in this case.

\subsection{Planetesimal formation}\label{sub:pf}

\begin{table}
\caption{Details on the planetesimal annuli: their final mass $M_{\rm plts}$ and radial extent (R$_{\rm in}$ and R$_{\rm out}$), formed in runs with different underlying protoplanetary disk models, global solids-to-gas ratio $Z$, turbulence strength $\alpha_{\rm t}$, and initial disk size R$_{\rm{disk}}$. }
\centering                         
\begin{tabular}{c l l l l l l}   
\hline\hline                
Disk\tablefootmark{a} & Z & $\alpha_{\rm t}$ & R$_{\rm{disk}}$\tablefootmark{b} & $M_{\rm plts}$\tablefootmark{c} & R$_{\rm in}$\tablefootmark{b} & R$_{\rm out}$\tablefootmark{b} \\   
\hline
          & 0.01 & $10^{-3}$ & 100 & --- & --- & --- \\
          & 0.02 & $10^{-3}$ & 100 & 3.54 & 1.98  & 2.34 \\
P-L & 0.03 & $10^{-3}$ & 100 & 24.28 & 1.93 & 2.99 \\
          & 0.04 & $10^{-3}$ & 100 & 62.76 & 1.89 & 3.90 \\
          & 0.05 & $10^{-3}$ & 100 & 115.10 & 1.87 & 4.90 \\
\hline
          & 0.03 & $10^{-4}$ & 100 & 217.13 & 1.66 & 5.19 \\
P-L & 0.03 & $3\cdot10^{-4}$ & 100 & 128.79 & 1.81 & 3.44 \\ 
          & 0.03 & $3\cdot10^{-3}$ & 100 & --- & --- & --- \\
          & 0.03 & $10^{-2}$ & 100 & --- & --- & --- \\  
\hline
          & 0.03 & $10^{-3}$ & 60 & 8.90 & 1.89 & 2.45 \\
P-L & 0.03 & $10^{-3}$ & 80 & 16.73 & 1.91 & 2.75 \\
          & 0.03 & $10^{-3}$ & 140 & 38.52 & 1.94 & 3.50 \\
          & 0.03 & $10^{-3}$ & 200 & 49.50 & 1.98 & 4.10 \\            
\hline
               & 0.01 & $10^{-3}$ & 100 & 119.80 & 1.03 & 2.20 \\
               & 0.02 & $10^{-3}$ & 100 & 371.37 & 0.98 & 4.08 \\
N-IRR & 0.03 & $10^{-3}$ & 100 & 663.18 & 0.96 & 5.50 \\
               & 0.04 & $10^{-3}$ & 100 & 928.35 & 0.96 & 6.83 \\
               & 0.05 & $10^{-3}$ & 100 & 1130.9 & 0.95 & 8.00 \\
\hline
            & 0.01 & $10^{-3}$ & 100 & --- & --- & --- \\
            & 0.02 & $10^{-3}$ & 100 & 3.32 & 2.12 & 4.33 \\
IRR  & 0.03 & $10^{-3}$ & 100 & 22.50 & 1.64 & 7.16 \\
            & 0.04 & $10^{-3}$ & 100 & 63.31 & 1.25 & 10.37  \\
            & 0.05 & $10^{-3}$ & 100 & 129.93 & 1.06 & 16.99 \\
\hline\hline
\end{tabular}
\tablefoot{\tablefoottext{a}{P-L -- power law, N-IRR -- non-irradiated, IRR - irradiated,} \tablefoottext{b}{in AU,} \tablefoottext{c}{in Earth masses}.}       
\label{table:allruns} 
\end{table}

Planetesimal formation is triggered in the pile-up arising outside of the snow line in many of the runs that we performed. Table~\ref{table:allruns} summarizes the information about mass and extent of the planetesimals annulus formed in models with different underlying gas parameters, initial dust-to-gas ratio, intrinsic turbulence level and disk extent. We notice that it is significantly easier to trigger planetesimal formation when the snow line is closer to the central star. The initial metallicity of $Z=0.01$ is sufficient for planetesimal formation in the non-irradiated disk model, which is the coldest (see the middle panel of Fig.~\ref{fig:disks}), while the power-law and the irradiated disks need at least $Z=0.02$. This is because close-in pile-up formation is aided by the process described by \citet{2016A&A...594A.105D}: the surface density of solids in the fragmentation dominated regime naturally evolves to $\Sigma_{\rm d}\propto r^{-3/2}$, a profile steeper than the gas disk, which leads to additional enhancement of dust-to-gas ratio in the inner disk. The further away the border between fragmentation dominated and drift dominated regions (see Fig.~\ref{fig:sizes}) is, the stronger enhancement we obtain. The maximum enhancement formed by this process would fall at the inner edge of the disk (although it is diffused in our models because the small aggregates inside of the snow line falling into the mixing regime). 
For this reason, the annuli formed closer to the central star tend to be more massive. 

The inner edge of the planetesimal formation zone depends primarily on the underlying disk model, and slightly changes with the assumed metallicity. This is because higher abundance of solids translates into higher flux of pebbles coming to the snow line, which deliver more water vapor and thus increase the vapor pressure that determines the snow line location (see Sect.~\ref{sub:evapcond}). At the same time, the higher the initial metallicity, the wider planetesimal annulus we get. The broadening of planetesimal annulus with higher incoming pebble flux is caused by the collective drift effect. The radial drift slows down near the solids-to-gas ratio peak and the more pebbles are delivered to this region, the wider this peak becomes. 

We stress that the collective drift effect is a critical component of our model without which obtaining the significant pile-up and planetesimal formation is nearly impossible. Recently, \citet{2017A&A...602A..21S} arrived at the same conclusion.  

\subsubsection{Impact of the turbulence strength}\label{sub:alpha}
  
  \begin{figure}
   \centering
   \includegraphics[width=0.9\hsize]{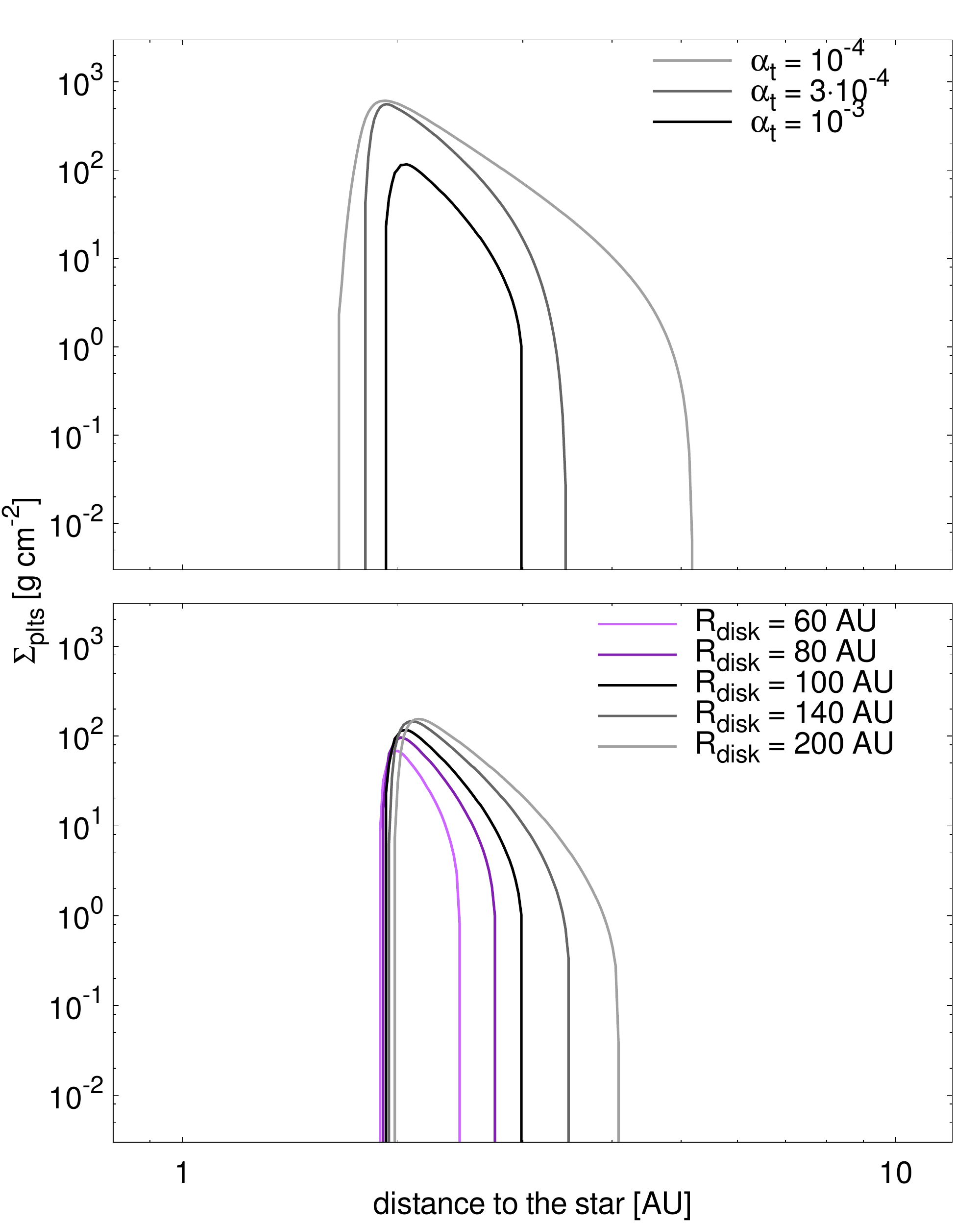}
      \caption{Planetesimals annuli obtained in the power-law disk models with different turbulence strength parameter $\alpha_{\rm t}$ (upper panel) and initial disk extent (bottom panel). Black solid line corresponds to the same model in both panels. }
      \label{fig:alpha_Rdisk}
\end{figure}

All the models presented in this paper up to this point were performed assuming that the fragmentation and settling of solids is regulated by turbulence with $\alpha_{\rm t}=10^{-3}$. 
However this value is rather vague, as it is very challenging to estimate the turbulence strength from the observational data. Recent estimates range from $\alpha_{\rm t}<10^{-3}$ for the outer parts of the disk around HD~163296 \citep{2015ApJ...813...99F} to $\alpha_{\rm t}\approx10^{-2}$ in the outer parts of the TW~Hya disk \citep{2016A&A...592A..49T}.

To test the impact of turbulence strength, we perform a suite of models where we vary the $\alpha_{\rm t}$ parameter value and keep all the other parameters constant. For this purpose, we use the setup with static, power-law disk and initial dust-to-gas ratio of $Z=0.03$, and vary the $\alpha_{\rm t}$ between $10^{-4}$ and $10^{-2}$. As can be seen in the upper panel of Fig.~\ref{fig:alpha_Rdisk}, lower values of the $\alpha_{\rm t}$ parameters facilitate planetesimal formation. The lower the turbulence strength, the wider and more massive the resulting planetesimal annulus becomes. We find that no planetesimal formation is possible for $\alpha_{\rm t}$ significantly higher than our fiducial $10^{-3}$ (see Table~\ref{table:allruns}). This is because the higher $\alpha_{\rm t}$ value reduces size of pebbles that can grow (see Eq.~(\ref{afrag})) and decreases the possibility of their settling (see Eq.~(\ref{hdust})), both of the factors counteract the possibility of obtaining conditions necessary to trigger the streaming instability, namely the minimum size of pebbles corresponding to ${\rm St}=10^{-2}$ and the midplane pebbles-to-gas ratio exceeding unity (see Sect.~\ref{sub:pfmodel}).

\subsubsection{Impact of the initial disk size}

The initial size of protoplanetary disk is rather uncertain. Observational constraints place the outer edge of the disk anywhere between 60~AU and several hundreds AU \citep{2009ApJ...700.1502A,2010ApJ...723.1241A}. Thus, we decided to test the impact that initial disk extent has on the planetesimal formation.

The bottom panel of Fig.~\ref{fig:alpha_Rdisk} shows the impact that initial distribution of material has. We used the power-law disk model with the initial dust-to-gas ratio of $Z=0.03$ and turbulence strength of $\alpha_{\rm t}=10^{-3}$ again and tested how the results change when the extent of this disk is different from fiducial 100~AU (however keeping the total mass of the disk constant). The variation in the resulting planetesimal formation is not quite as pronounced as when varying the turbulence strength, but the larger the disk, the more massive and more extended the planetesimals annulus becomes. This is because larger disks provide long lasting supply of pebbles, as their growth takes longer at larger orbital distances. In other words, increasing the disk size (while keeping its mass unchanged) shifts more solid mass to its outer regions, and this reservoir can be then used to form more planetesimals.

\subsubsection{Pebbles and planetesimals composition}

  \begin{figure}
   \centering
   \includegraphics[width=0.9\hsize]{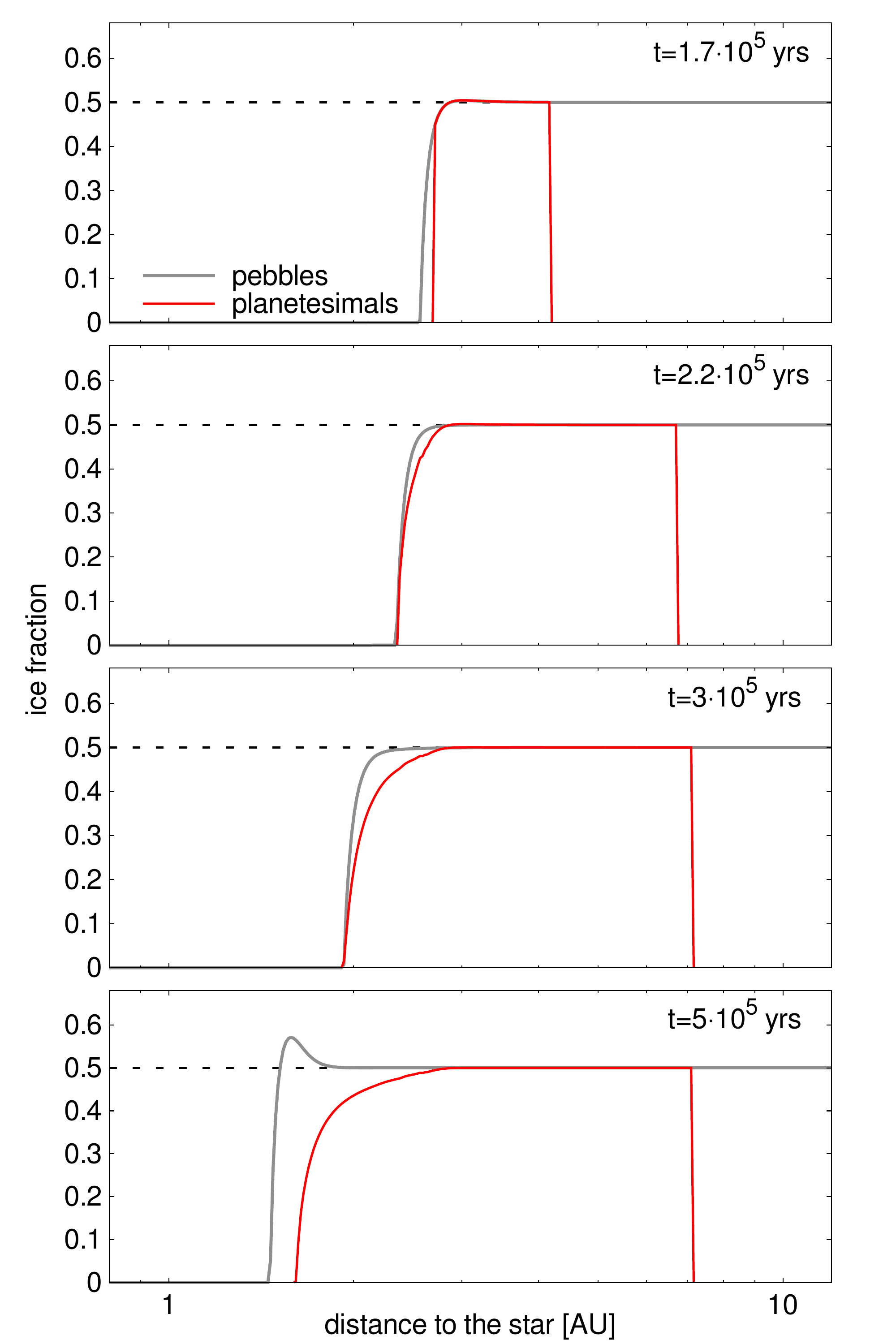}
      \caption{Panels show time evolution of the ice fraction of planetesimals (red) and pebbles (gray) for the irradiated disk model with initial solids-to-gas ratio of 0.03. The initial ice mass fraction (outside of the snow line) of 0.5 is marked with the dashed line.}
      \label{fig:iceratio}
\end{figure}

Figure~\ref{fig:iceratio} presents the time evolution of ice fraction of pebbles (gray line) and ice fraction of the resulting planetesimals (red line) for the irradiated disk model with initial metallicity of 0.03 (the same model as presented on Figs.~\ref{fig:7panels} and \ref{fig:sizes}). The first panel is plotted at the beginning of planetesimal formation stage and the last panel corresponds to a time shortly after planetesimal formation is terminated. We assume that the composition of planetesimals reflects those of pebbles from which they are forming.
We start all our models with dust outside of the snow line consisting of 50\% water ice (dashed line). Evaporation removes the solid ice that is delivered to the inner disk by radial drift and turns it to vapor, part of which is able to diffuse outwards and recondense, moderately enhancing the ice content of dust aggregates just outside of the snow line. As explained in Sect.~\ref{sub:what}, the main cause for the pile-up of pebbles outside of the snow line is the "traffic jam" arising in the inner disk and not recondensation, which is reflected here in the low amplitude of the ice fraction enhancement. When planetesimal formation starts after about $1.7\cdot10^5$~years of evolution, the enhancement of pebbles ice fraction goes away completely. This is because the streaming instability turns the icy pebbles from outside of the snow line to planetesimals and thus hinders delivery of water to the evaporation region, and decreases the rate of recondensation. During the planetesimal formation phase, the snow line moves inwards as the disk evolves and thus the ice content of the final planetesimal population decreases more smoothly than those of pebbles, which are evaporating rather quickly. After the planetesimal formation is over ($\sim4\cdot10^5$~years of evolution), pebbles composition is set again by the interplay of evaporation and recondensation, causing the mild enhancement of ice fraction outside of the evaporation region.

\section{Discussion}\label{sub:discussion}

\subsection{Comparison to published work}

\citet{2016ApJ...828L...2A} presented similar idea where pebbles drift radially and pile-up to reach conditions required for planetesimal formation. However, they neglected dust growth and fragmentation, assuming constant pebble size throughout the disk. In such a setup, the Stokes number increases with the radial distance (because the gas surface density drops, see Eq.~(\ref{stnr})), making it easier to trigger planetesimal formation further away from the star. At the same time, the particles further away in the disk drift faster (because the radial drift speed depends on the Stokes number, not size, see Eq.~(\ref{vbar})), which leads to a pile-up that is harder to generate when particle sizes are decided by fragmentation and radial drift. As derived by \citet{2012A&A...539A.148B}, in the fragmentation-dominated regime the steady-state dust surface density is proportional to $r^{-1.5}$ and in the radial drift regime to $r^{-0.75}$. Analogous derivation assuming constant dust size gives $\Sigma_{\rm}\propto r^{-2}$, which facilitates obtaining high dust-to-gas ratio in the inner disk more than our models.

Recently, \citet{2017A&A...602A..21S} performed local models focusing on the snow line region. They took into account the inflow of icy pebbles that would be formed in the outer disk and brought to the snow line region by radial drift and the outflow of water vapor carried with gas accretion. They found that water diffusion and recondensation enhances surface density of icy pebbles by a factor of 3-5 outside the snow line that is further increased if the evaporating pebbles release many small refractory "seeds" that help the pile-up thanks to the "traffic jam" effect and outward diffusion. With the models presented in this paper, although they are fundamentally different by construction, as \citet{2017A&A...602A..21S} focused on a local box and used particle approach, while in this paper we perform global disk models applying fluid approach to dust dynamics, we find very similar both qualitative and quantitative results. This is very encouraging and proving that a pile-up of pebbles outside the snow line is a robust mechanism that can trigger formation of the first, icy planetesimals.

On the other hand, another recent work by \citet{2016A&A...596L...3I} suggested different scenario of planetesimal formation, namely pile-up of dry dust grains released by the icy pebbles inside of the snow line. They find that for a sufficiently high flux of pebbles, the dust-to-gas ratio increases to a point when direct gravitational instability is possible. We do also find an increase in the solids-to-gas ratio inside of the snow line but it is not as significant. The main difference is that we assume that the small grains released from the icy pebbles quickly coagulate until they reach fragmentation limit at about millimeter sizes and that they are vertically mixed by the turbulence, while \citet{2016A&A...596L...3I} consider that the grains stay at micron sized and that their scale-height is equal to the scale-height of icy pebbles that released them. 

Both \citet{2016A&A...596L...3I} and \citet{2017A&A...602A..21S} stressed the importance of sufficiently high pebble mass flux for the possibility of planetesimal formation. In their models, the pebble flux was a free parameter since they did not include dust growth to pebble sizes and their drift self-consistently. With our models, we can measure the pebble flux incoming to the snow line region. For the power-law disk model, which is similar to the models used in the quoted papers, we measure the ratio of pebble mass flux to the gas mass flux during the planetesimal formation stage on the order of 0.5, which \citet{2017A&A...602A..21S} also found sufficient for triggering planetesimal formation via streaming instability outside of the snow line. 

\subsection{Implications for planet formation}

Our results suggest that formation of planetesimals is simpler in the direct vicinity of the water evaporation front.
This could naturally explain the fast formation of Jupiter in the Solar System \citep{2017LPI....48.1386K}. 
The surface density obtained in the planetesimal annulus in all of our runs is higher than predicted by the minimum mass solar nebula models \citep{1977Ap&SS..51..153W, 1981PThPS..70...35H}. Starting from a high mass concentration translates into faster growth of planetesimals to planetary embryos and the planetary cores. It is known that enhancement of about 10~times over the solids surface density corresponding to the minimum mass solar nebula is necessary to allow for Jupiter core formation before the gas disk dispersal \citep{1996Icar..124...62P, 2000ApJ...537.1013I, 2010Icar..209..836K}. 
The fast formation of a gas giant outside of the snow line could possibly halt the delivery of water to the inner part of planetary system \citep{2015Icar..258..418M}. If such a barrier is not formed quickly, it may be problematic to explain the low water content of terrestrial planets in the Solar System \citep{2016A&A...589A..15S}.

Our results, including the self-consistent surface densities and composition of planetesimals and pebbles, may be used as an input to models dealing with later stages of planet accretion, particularly discussing the pebble accretion process, when the planetary cores grow by accreting not only planetesimals, but also the leftover pebbles that were not incorporated during the planetesimal formation stage. The radial dependence of sizes and radial flux of pebbles that we can extract from our results are important parameters of the pebble accretion models \citep{2010A&A...520A..43O, 2014A&A...572A.107L, 2015Natur.524..322L, 2016A&A...586A..66V}.

One prominent consequence of the planetesimal formation mechanism we discuss is that the first planetesimals are water-rich (see Fig.~\ref{fig:iceratio}). 
As a consequence, planetary cores formed from these planetesimals would also be water-rich. For small mass planets (without a massive gas 
envelope), the presence of large amounts of water may be detrimental for habitability (see \citet{2013A&A...558A.109A, 2015MNRAS.452.3752K}, see however \citet{2017ApJ...838...24L} for another view). Planetesimal formation mechanism we describe here could therefore imply that the majority 
of low mass planets are not habitable. However, the recent models show that low mass, short period planets detected by the {\it Kepler} mission should be water-poor \citep[see e.g.][]{2017arXiv170600251J}. In the framework of our planetesimal formation model, this imply that other mechanisms, facilitating efficient water loss from existing planetesimals or allowing the formation of dry planetesimals, are at work to prevent the accumulation of water on these short-period planets.

\section{Summary}\label{sub:summary}

In this paper, we addressed the problem of connection between dust evolution and planetesimal formation, which is still one of the least certain aspects of the planet formation theory. As dust growth is hindered by collisional fragmentation and radial drift, continuous growth from micron to planetesimal sizes appears to be improbable. We propose that the first planetesimals form via streaming instability, in a pile-up of icy pebbles generated outside of the snow line. 

The water snow line is a favorable location for planetesimal formation as large icy pebbles efficiently deliver water and embedded refractory material to the inner part of the disk. The water vapor is partially mixed outwards by diffusion and recondenses just outside of the snow line, locally enhancing the solids-to-gas ratio. At the same time, the less sticky, dry aggregates inside of the snow line drift at much lower speed, creating the "traffic jam" effect and helping to reach high dust concentration which is needed to form planetesimals in the streaming instability scenario. 

A relatively compact annulus of icy planetesimal is a common result of our simulations, performed with three diverse protoplanetary disk models and different input parameters. The main condition we find for making this outcome feasible is that the turbulence strength cannot be too high: the corresponding $\alpha_{\rm t}$ parameter must stay equal to or below $10^{-3}$. Also, the further away the snow line is located, the higher disk metallicity is needed to allow for planetesimal formation.

On a more general level, this work as well as similar studies \citep{2016A&A...594A.105D, 2017ApJ...839...16C, 2017MNRAS.467.1984G} indicate that dust distribution during and after the planetesimal formation stage is very different from the commonly assumed power-laws, as significant redistribution of solids must take place before the conditions necessary for planetesimal formation happen. In particular, the solids-to-gas ratio is significantly increased at the location where planetesimals form, which should facilitate faster accretion of the final planets.

\begin{acknowledgements}
     We thank the anonymous referee for their detailed reports that helped us to improve this paper. This work has been carried out within the framework of the National Centre for Competence in Research PlanetS supported by the Swiss National Science Foundation. The authors acknowledge the financial support of the SNSF.
\end{acknowledgements}

\bibliographystyle{aa} 
\bibliography{snowline.bib}

\begin{thebibliography}{70}
\expandafter\ifx\csname natexlab\endcsname\relax\def\natexlab#1{#1}\fi

\bibitem[{{Alibert} {et~al.}(2013){Alibert}, {Carron}, {Fortier}, {Pfyffer},
  {Benz}, {Mordasini}, \& {Swoboda}}]{2013A&A...558A.109A}
{Alibert}, Y., {Carron}, F., {Fortier}, A., {et~al.} 2013, \aap, 558, A109

\bibitem[{{Alibert} {et~al.}(2005){Alibert}, {Mordasini}, {Benz}, \&
  {Winisdoerffer}}]{2005A&A...434..343A}
{Alibert}, Y., {Mordasini}, C., {Benz}, W., \& {Winisdoerffer}, C. 2005, \aap,
  434, 343

\bibitem[{{Andrews} {et~al.}(2009){Andrews}, {Wilner}, {Hughes}, {Qi}, \&
  {Dullemond}}]{2009ApJ...700.1502A}
{Andrews}, S.~M., {Wilner}, D.~J., {Hughes}, A.~M., {Qi}, C., \& {Dullemond},
  C.~P. 2009, \apj, 700, 1502

\bibitem[{{Andrews} {et~al.}(2010){Andrews}, {Wilner}, {Hughes}, {Qi}, \&
  {Dullemond}}]{2010ApJ...723.1241A}
{Andrews}, S.~M., {Wilner}, D.~J., {Hughes}, A.~M., {Qi}, C., \& {Dullemond},
  C.~P. 2010, \apj, 723, 1241

\bibitem[{{Armitage} {et~al.}(2016){Armitage}, {Eisner}, \&
  {Simon}}]{2016ApJ...828L...2A}
{Armitage}, P.~J., {Eisner}, J.~A., \& {Simon}, J.~B. 2016, \apjl, 828, L2

\bibitem[{{Aumatell} \& {Wurm}(2014)}]{2014MNRAS.437..690A}
{Aumatell}, G. \& {Wurm}, G. 2014, \mnras, 437, 690

\bibitem[{{Bai}(2016)}]{2016ApJ...821...80B}
{Bai}, X.-N. 2016, \apj, 821, 80

\bibitem[{{Bai} \& {Stone}(2010)}]{2010ApJ...722.1437B}
{Bai}, X.-N. \& {Stone}, J.~M. 2010, \apj, 722, 1437

\bibitem[{{Banzatti} {et~al.}(2015){Banzatti}, {Pinilla}, {Ricci},
  {Pontoppidan}, {Birnstiel}, \& {Ciesla}}]{2015ApJ...815L..15B}
{Banzatti}, A., {Pinilla}, P., {Ricci}, L., {et~al.} 2015, \apjl, 815, L15

\bibitem[{{Birnstiel} {et~al.}(2010){Birnstiel}, {Dullemond}, \&
  {Brauer}}]{2010A&A...513A..79B}
{Birnstiel}, T., {Dullemond}, C.~P., \& {Brauer}, F. 2010, \aap, 513, A79

\bibitem[{{Birnstiel} {et~al.}(2012){Birnstiel}, {Klahr}, \&
  {Ercolano}}]{2012A&A...539A.148B}
{Birnstiel}, T., {Klahr}, H., \& {Ercolano}, B. 2012, \aap, 539, A148

\bibitem[{{Bitsch} {et~al.}(2015){Bitsch}, {Johansen}, {Lambrechts}, \&
  {Morbidelli}}]{2015A&A...575A..28B}
{Bitsch}, B., {Johansen}, A., {Lambrechts}, M., \& {Morbidelli}, A. 2015, \aap,
  575, A28

\bibitem[{{Brauer} {et~al.}(2008){Brauer}, {Henning}, \&
  {Dullemond}}]{2008A&A...487L...1B}
{Brauer}, F., {Henning}, T., \& {Dullemond}, C.~P. 2008, \aap, 487, L1

\bibitem[{{Carrera} {et~al.}(2017){Carrera}, {Gorti}, {Johansen}, \&
  {Davies}}]{2017ApJ...839...16C}
{Carrera}, D., {Gorti}, U., {Johansen}, A., \& {Davies}, M.~B. 2017, \apj, 839,
  16

\bibitem[{{Carrera} {et~al.}(2015){Carrera}, {Johansen}, \&
  {Davies}}]{2015A&A...579A..43C}
{Carrera}, D., {Johansen}, A., \& {Davies}, M.~B. 2015, \aap, 579, A43

\bibitem[{{Ciesla} \& {Cuzzi}(2006)}]{2006Icar..181..178C}
{Ciesla}, F.~J. \& {Cuzzi}, J.~N. 2006, \icarus, 181, 178

\bibitem[{{Cridland} {et~al.}(2017){Cridland}, {Pudritz}, \&
  {Birnstiel}}]{2017MNRAS.465.3865C}
{Cridland}, A.~J., {Pudritz}, R.~E., \& {Birnstiel}, T. 2017, \mnras, 465, 3865

\bibitem[{{Cuzzi} \& {Zahnle}(2004)}]{2004ApJ...614..490C}
{Cuzzi}, J.~N. \& {Zahnle}, K.~J. 2004, \apj, 614, 490

\bibitem[{{Dr{\c a}{\.z}kowska} {et~al.}(2016){Dr{\c a}{\.z}kowska}, {Alibert},
  \& {Moore}}]{2016A&A...594A.105D}
{Dr{\c a}{\.z}kowska}, J., {Alibert}, Y., \& {Moore}, B. 2016, \aap, 594, A105

\bibitem[{{Dr{\c a}{\.z}kowska} \& {Dullemond}(2014)}]{2014A&A...572A..78D}
{Dr{\c a}{\.z}kowska}, J. \& {Dullemond}, C.~P. 2014, \aap, 572, A78

\bibitem[{{Dr{\c a}{\.z}kowska} {et~al.}(2013){Dr{\c a}{\.z}kowska},
  {Windmark}, \& {Dullemond}}]{2013A&A...556A..37D}
{Dr{\c a}{\.z}kowska}, J., {Windmark}, F., \& {Dullemond}, C.~P. 2013, \aap,
  556, A37

\bibitem[{{Dubrulle} {et~al.}(1995){Dubrulle}, {Morfill}, \&
  {Sterzik}}]{1995Icar..114..237D}
{Dubrulle}, B., {Morfill}, G., \& {Sterzik}, M. 1995, \icarus, 114, 237

\bibitem[{{Dzyurkevich} {et~al.}(2013){Dzyurkevich}, {Turner}, {Henning}, \&
  {Kley}}]{2013ApJ...765..114D}
{Dzyurkevich}, N., {Turner}, N.~J., {Henning}, T., \& {Kley}, W. 2013, \apj,
  765, 114

\bibitem[{{Estrada} {et~al.}(2016){Estrada}, {Cuzzi}, \&
  {Morgan}}]{2016ApJ...818..200E}
{Estrada}, P.~R., {Cuzzi}, J.~N., \& {Morgan}, D.~A. 2016, \apj, 818, 200

\bibitem[{{Flaherty} {et~al.}(2015){Flaherty}, {Hughes}, {Rosenfeld},
  {Andrews}, {Chiang}, {Simon}, {Kerzner}, \& {Wilner}}]{2015ApJ...813...99F}
{Flaherty}, K.~M., {Hughes}, A.~M., {Rosenfeld}, K.~A., {et~al.} 2015, \apj,
  813, 99

\bibitem[{{Gonzalez} {et~al.}(2017){Gonzalez}, {Laibe}, \&
  {Maddison}}]{2017MNRAS.467.1984G}
{Gonzalez}, J.-F., {Laibe}, G., \& {Maddison}, S.~T. 2017, \mnras, 467, 1984

\bibitem[{{Gundlach} \& {Blum}(2015)}]{2015ApJ...798...34G}
{Gundlach}, B. \& {Blum}, J. 2015, \apj, 798, 34

\bibitem[{{Gundlach} {et~al.}(2011){Gundlach}, {Kilias}, {Beitz}, \&
  {Blum}}]{2011Icar..214..717G}
{Gundlach}, B., {Kilias}, S., {Beitz}, E., \& {Blum}, J. 2011, \icarus, 214,
  717

\bibitem[{{G{\"u}ttler} {et~al.}(2010){G{\"u}ttler}, {Blum}, {Zsom}, {Ormel},
  \& {Dullemond}}]{2010A&A...513A..56G}
{G{\"u}ttler}, C., {Blum}, J., {Zsom}, A., {Ormel}, C.~W., \& {Dullemond},
  C.~P. 2010, \aap, 513, A56

\bibitem[{{Hartmann} {et~al.}(1998){Hartmann}, {Calvet}, {Gullbring}, \&
  {D'Alessio}}]{1998ApJ...495..385H}
{Hartmann}, L., {Calvet}, N., {Gullbring}, E., \& {D'Alessio}, P. 1998, \apj,
  495, 385

\bibitem[{{Hayashi}(1981)}]{1981PThPS..70...35H}
{Hayashi}, C. 1981, Progress of Theoretical Physics Supplement, 70, 35

\bibitem[{{Hughes} \& {Armitage}(2012)}]{2012MNRAS.423..389H}
{Hughes}, A.~L.~H. \& {Armitage}, P.~J. 2012, \mnras, 423, 389

\bibitem[{{Ida} \& {Guillot}(2016)}]{2016A&A...596L...3I}
{Ida}, S. \& {Guillot}, T. 2016, \aap, 596, L3

\bibitem[{{Ikoma} {et~al.}(2000){Ikoma}, {Nakazawa}, \&
  {Emori}}]{2000ApJ...537.1013I}
{Ikoma}, M., {Nakazawa}, K., \& {Emori}, H. 2000, \apj, 537, 1013

\bibitem[{{Jin} \& {Mordasini}(2017)}]{2017arXiv170600251J}
{Jin}, S. \& {Mordasini}, C. 2017, ArXiv e-prints [\eprint[arXiv]{1706.00251}]

\bibitem[{{Johansen} {et~al.}(2011){Johansen}, {Klahr}, \&
  {Henning}}]{2011A&A...529A..62J}
{Johansen}, A., {Klahr}, H., \& {Henning}, T. 2011, \aap, 529, A62

\bibitem[{{Johansen} {et~al.}(2007){Johansen}, {Oishi}, {Mac Low}, {Klahr},
  {Henning}, \& {Youdin}}]{2007Natur.448.1022J}
{Johansen}, A., {Oishi}, J.~S., {Mac Low}, M.-M., {et~al.} 2007, \nat, 448,
  1022

\bibitem[{{Kanagawa} {et~al.}(2017){Kanagawa}, {Ueda}, {Muto}, \&
  {Okuzumi}}]{2017ApJ...844..142K}
{Kanagawa}, K.~D., {Ueda}, T., {Muto}, T., \& {Okuzumi}, S. 2017, \apj, 844,
  142

\bibitem[{{Kataoka} {et~al.}(2013){Kataoka}, {Tanaka}, {Okuzumi}, \&
  {Wada}}]{2013A&A...557L...4K}
{Kataoka}, A., {Tanaka}, H., {Okuzumi}, S., \& {Wada}, K. 2013, \aap, 557, L4

\bibitem[{{Kitzmann} {et~al.}(2015){Kitzmann}, {Alibert}, {Godolt}, {Grenfell},
  {Heng}, {Patzer}, {Rauer}, {Stracke}, \& {von Paris}}]{2015MNRAS.452.3752K}
{Kitzmann}, D., {Alibert}, Y., {Godolt}, M., {et~al.} 2015, \mnras, 452, 3752

\bibitem[{{Kobayashi} {et~al.}(2010){Kobayashi}, {Tanaka}, {Krivov}, \&
  {Inaba}}]{2010Icar..209..836K}
{Kobayashi}, H., {Tanaka}, H., {Krivov}, A.~V., \& {Inaba}, S. 2010, \icarus,
  209, 836

\bibitem[{{Kowalik} {et~al.}(2013){Kowalik}, {Hanasz}, {W{\'o}lta{\'n}ski}, \&
  {Gawryszczak}}]{2013MNRAS.434.1460K}
{Kowalik}, K., {Hanasz}, M., {W{\'o}lta{\'n}ski}, D., \& {Gawryszczak}, A.
  2013, \mnras, 434, 1460

\bibitem[{{Kretke} \& {Lin}(2007)}]{2007ApJ...664L..55K}
{Kretke}, K.~A. \& {Lin}, D.~N.~C. 2007, \apjl, 664, L55

\bibitem[{{Krijt} {et~al.}(2016{\natexlab{a}}){Krijt}, {Ciesla}, \&
  {Bergin}}]{2016ApJ...833..285K}
{Krijt}, S., {Ciesla}, F.~J., \& {Bergin}, E.~A. 2016{\natexlab{a}}, \apj, 833,
  285

\bibitem[{{Krijt} {et~al.}(2016{\natexlab{b}}){Krijt}, {Ormel}, {Dominik}, \&
  {Tielens}}]{2016A&A...586A..20K}
{Krijt}, S., {Ormel}, C.~W., {Dominik}, C., \& {Tielens}, A.~G.~G.~M.
  2016{\natexlab{b}}, \aap, 586, A20

\bibitem[{{Kruijer} {et~al.}(2017){Kruijer}, {Kleine}, {Burkhardt}, \&
  {Budde}}]{2017LPI....48.1386K}
{Kruijer}, T.~S., {Kleine}, T., {Burkhardt}, C., \& {Budde}, G. 2017, in Lunar
  and Planetary Inst.~Technical Report, Vol.~48, Lunar and Planetary Science
  Conference, 1386

\bibitem[{{Lambrechts} \& {Johansen}(2014)}]{2014A&A...572A.107L}
{Lambrechts}, M. \& {Johansen}, A. 2014, \aap, 572, A107

\bibitem[{{Levi} {et~al.}(2017){Levi}, {Sasselov}, \&
  {Podolak}}]{2017ApJ...838...24L}
{Levi}, A., {Sasselov}, D., \& {Podolak}, M. 2017, \apj, 838, 24

\bibitem[{{Levison} {et~al.}(2015){Levison}, {Kretke}, \&
  {Duncan}}]{2015Natur.524..322L}
{Levison}, H.~F., {Kretke}, K.~A., \& {Duncan}, M.~J. 2015, \nat, 524, 322

\bibitem[{{Lichtenegger} \& {Komle}(1991)}]{1991Icar...90..319L}
{Lichtenegger}, H.~I.~M. \& {Komle}, N.~I. 1991, \icarus, 90, 319

\bibitem[{{Morbidelli} {et~al.}(2015){Morbidelli}, {Lambrechts}, {Jacobson}, \&
  {Bitsch}}]{2015Icar..258..418M}
{Morbidelli}, A., {Lambrechts}, M., {Jacobson}, S., \& {Bitsch}, B. 2015,
  \icarus, 258, 418

\bibitem[{{Okuzumi} {et~al.}(2012){Okuzumi}, {Tanaka}, {Kobayashi}, \&
  {Wada}}]{2012ApJ...752..106O}
{Okuzumi}, S., {Tanaka}, H., {Kobayashi}, H., \& {Wada}, K. 2012, \apj, 752,
  106

\bibitem[{{Ormel} \& {Klahr}(2010)}]{2010A&A...520A..43O}
{Ormel}, C.~W. \& {Klahr}, H.~H. 2010, \aap, 520, A43

\bibitem[{{Pollack} {et~al.}(1996){Pollack}, {Hubickyj}, {Bodenheimer},
  {Lissauer}, {Podolak}, \& {Greenzweig}}]{1996Icar..124...62P}
{Pollack}, J.~B., {Hubickyj}, O., {Bodenheimer}, P., {et~al.} 1996, \icarus,
  124, 62

\bibitem[{{Ros} \& {Johansen}(2013)}]{2013A&A...552A.137R}
{Ros}, K. \& {Johansen}, A. 2013, \aap, 552, A137

\bibitem[{{Sato} {et~al.}(2016){Sato}, {Okuzumi}, \&
  {Ida}}]{2016A&A...589A..15S}
{Sato}, T., {Okuzumi}, S., \& {Ida}, S. 2016, \aap, 589, A15

\bibitem[{{Schoonenberg} \& {Ormel}(2017)}]{2017A&A...602A..21S}
{Schoonenberg}, D. \& {Ormel}, C.~W. 2017, \aap, 602, A21

\bibitem[{{Shakura} \& {Sunyaev}(1973)}]{1973A&A....24..337S}
{Shakura}, N.~I. \& {Sunyaev}, R.~A. 1973, \aap, 24, 337

\bibitem[{{Simon} {et~al.}(2016){Simon}, {Armitage}, {Li}, \&
  {Youdin}}]{2016ApJ...822...55S}
{Simon}, J.~B., {Armitage}, P.~J., {Li}, R., \& {Youdin}, A.~N. 2016, \apj,
  822, 55

\bibitem[{{Stammler} {et~al.}(2017){Stammler}, {Birnstiel}, {Pani{\'c}},
  {Dullemond}, \& {Dominik}}]{2017A&A...600A.140S}
{Stammler}, S.~M., {Birnstiel}, T., {Pani{\'c}}, O., {Dullemond}, C.~P., \&
  {Dominik}, C. 2017, \aap, 600, A140

\bibitem[{{Stevenson} \& {Lunine}(1988)}]{1988Icar...75..146S}
{Stevenson}, D.~J. \& {Lunine}, J.~I. 1988, \icarus, 75, 146

\bibitem[{{Teague} {et~al.}(2016){Teague}, {Guilloteau}, {Semenov}, {Henning},
  {Dutrey}, {Pi{\'e}tu}, {Birnstiel}, {Chapillon}, {Hollenbach}, \&
  {Gorti}}]{2016A&A...592A..49T}
{Teague}, R., {Guilloteau}, S., {Semenov}, D., {et~al.} 2016, \aap, 592, A49

\bibitem[{{Turner} {et~al.}(2014){Turner}, {Fromang}, {Gammie}, {Klahr},
  {Lesur}, {Wardle}, \& {Bai}}]{2014prpl.conf..411T}
{Turner}, N.~J., {Fromang}, S., {Gammie}, C., {et~al.} 2014, Protostars and
  Planets VI, 411

\bibitem[{{Visser} \& {Ormel}(2016)}]{2016A&A...586A..66V}
{Visser}, R.~G. \& {Ormel}, C.~W. 2016, \aap, 586, A66

\bibitem[{{Wada} {et~al.}(2011){Wada}, {Tanaka}, {Suyama}, {Kimura}, \&
  {Yamamoto}}]{2011ApJ...737...36W}
{Wada}, K., {Tanaka}, H., {Suyama}, T., {Kimura}, H., \& {Yamamoto}, T. 2011,
  \apj, 737, 36

\bibitem[{{Wang}(2015)}]{2015MNRAS.449.1084W}
{Wang}, X.-M. 2015, \mnras, 449, 1084

\bibitem[{{Weidenschilling}(1977)}]{1977Ap&SS..51..153W}
{Weidenschilling}, S.~J. 1977, \apss, 51, 153

\bibitem[{{Wuchterl} {et~al.}(2000){Wuchterl}, {Guillot}, \&
  {Lissauer}}]{2000prpl.conf.1081W}
{Wuchterl}, G., {Guillot}, T., \& {Lissauer}, J.~J. 2000, Protostars and
  Planets IV, 1081

\bibitem[{{Yang} {et~al.}(2016){Yang}, {Johansen}, \&
  {Carrera}}]{2016arXiv161107014Y}
{Yang}, C.-C., {Johansen}, A., \& {Carrera}, D. 2016, ArXiv e-prints
  [\eprint[arXiv]{1611.07014}]

\bibitem[{{Youdin} \& {Goodman}(2005)}]{2005ApJ...620..459Y}
{Youdin}, A.~N. \& {Goodman}, J. 2005, \apj, 620, 459

\end{thebibliography}

\end{document}